\documentstyle[12pt,graphics]{article}

\title{Modelling the Pioneer anomaly \\ as modified inertia}

\author{M.E.~McCulloch \footnote{Met Office, FitzRoy Road, Exeter, EX1 3PB, UK.}}

\begin{document}

\maketitle

\section*{ABSTRACT}

This paper proposes an explanation for the Pioneer anomaly: an unexplained Sunward 
acceleration of 8.74 $\pm$ 1.33 $\times$ 10$^{-10}$ m~s$^{-2}$ seen in the behaviour 
of the Pioneer probes. Two hypotheses are made: (1) Inertia is a reaction to Unruh 
radiation and (2) this reaction is weaker for low accelerations because some 
wavelengths in the Unruh spectrum do not fit within a limiting scale (twice the
Hubble distance) and are disallowed: a process similar to the Casimir effect. 
When these ideas are used to model the Pioneer crafts' trajectories there is a 
slight reduction in their inertial mass, causing an anomalous Sunward acceleration 
of 6.9 $\pm$ 3.5 $\times$ 10$^{-10}$ m~s$^{-2}$ which agrees within error bars with 
the observed Pioneer anomaly beyond 10 AU from the Sun. This new scheme is appealingly 
simple and does not require adjustable parameters. However, it also predicts an anomaly 
within 10 AU of the Sun, which has not been observed. Various observational tests 
for the idea are proposed.

\newpage

\section{INTRODUCTION}

Anderson~$et~al.$~(1998) have detected a constant unexplained acceleration of both 
Pioneer 10 and 11 of 8.74 $\pm$ 1.33 $\times$ 10$^{-10}$ m~s$^{-2}$ directed approximately 
towards the Sun. Since the behaviour of the Pioneer craft should be predictable 
because of their spin-stabilisation (Anderson~$et~al.$~1998,~2002) but disagrees 
with our present understanding of motion, and since no convincing mundane physical 
explanation has so far been successful, the anomaly will be assumed here to be real.

Combining Newton's second law, and his law of gravity, the acceleration of a body 
of gravitational mass $m_g$ due to a larger body of mass $M$ at a distance $r$ is
\begin{equation}
a = \frac{GM m_g}{m_i r^2} \label{eq:n2ndlaw},
\end{equation}
where $m_i$ is the inertial mass and $G$ is Newton's gravity constant. Usually we assume 
that $m_i=m_g$ (the equivalence principle). However, this formula shows that to account 
for the anomalous acceleration $a$ of the Pioneer craft towards the Sun we can 
increase $G$, increase $M$, or increase $m_g/m_i$.

The Pioneer anomaly is similar to the galaxy rotation problem which also involves an 
unexplained acceleration towards a centre of mass. One solution to this problem was 
proposed by Milgrom~(1983) and is called MOdified Newtonian Dynamics (MOND). This 
theory has proved successful in reproducing galaxy rotation curves and is usually 
(but not necessarily) based on the first approach mentioned above: $G$ is increased 
for accelerations lower than 1.2 $\times$ $10^{-10}~m~s^{-2}$. 
This is also the approach of the relativistic extension of MOND 
by Bekenstein~(2004) which is called TeVeS. As an alternative, $G$ can be modified at long 
distances. This is the approach taken by the STVG theory of Moffat and Brownstein~(2006) 
which has been used to model the Pioneer anomaly, though they need adjustable parameters 
to do this. The conformal gravity theory of Mannheim~(1990) also modifies $G$ so that it 
is repulsive at long distances. 

An example of the second approach (increasing $M$) is the dark matter hypothesis of 
Zwicky (1933). Excess, invisible, matter is added to the galaxy to explain the 
implied extra centripetal acceleration. However dark matter fits to galaxies 
have three free parameters, whereas MOND has only one: the mass to light ratio 
(Sellwood,~2004).

The third approach, reducing the inertial mass ($m_i$), was first suggested by
Milgrom~(1983) who realised that MOND could be explained as a modification of
inertia instead of $G$. In later papers (Milgrom,~1994,~1999) he suggested
a possible physical cause for the inertial version of MOND which is discussed
in section 2.1 below. As he noted, there are 
some observations that imply that it is inertia that should be modified and 
not $G$ or $M$. For example: the possible change in behaviour of the Pioneer craft 
upon moving from a bound to an unbound trajectory (to be confirmed, or not, soon, 
by the Pioneer team), and the planets, which are on bound orbits, do not seem 
to show the anomaly. Also, MOND behaviour in galaxies begins below a limiting 
acceleration and not beyond a limiting distance, as noticed by Saunders 
and McGaugh~(2002).

One possibility for a model of inertia is that of Haisch~$et~al.$~(1994) who proposed 
that an accelerated object feels a magnetic Lorentz force through its interaction with 
a zero point field (ZPF) similar to the Unruh field (Unruh,~1976). This force is given 
by $F=-{\Gamma \omega_c^2 \hbar a}/{2 \pi c^2}$ 
where $\Gamma$ is the Abraham-Lorentz damping constant of the parton being 
oscillated, $\hbar$ is the reduced Planck constant, $\omega_c$ is the 
Compton scale of the parton below which the oscillations of the ZPF have no 
effect on it, $c$ is the speed of light, and $a$ is acceleration. 
Haisch~$et~al.$~(1994) showed that this force behaves like inertia.

One objection to a modification of inertia is that it violates the equivalence principle, 
which has recently been tested to an accuracy of $10^{-13}$~kg by Baessler~$et~al$~(1999).
However, this principle has not been tested at the low accelerations seen by the Pioneer 
craft or by stars at the edges of galaxies.

\section{THE MODEL}

\subsection{Unruh radiation curtailed at the Hubble distance.}

After work by Hawking~(1974), Unruh~(1976) showed that a body with an acceleration $a$ 
sees thermal radiation of temperature $T$ where
    \begin{equation}
    T = \frac{\hbar a}{2 \pi ck},     \label{eq:unruh}
    \end{equation}
where $k$ is Boltzmann's constant. The dominant wavelength of this radiation 
($\lambda_m$) is given by Wien's displacement law ($\lambda_m=W/T$), where $W$
is Wien's constant. Replacing $T$ using (\ref{eq:unruh}) and $W$ with $\beta h c / k$, 
where $\beta=0.2$ leaves
    \begin{equation} 
    \lambda_m = \frac{4 \pi^2 \beta c^2}{a}.    \label{eq:wiens}
    \end{equation}
Milgrom~(1994,~1999) realised that as the acceleration decreases the wavelength $\lambda_m$ 
increases, and eventually becomes as large as the Hubble distance ($c/H$) where $H$ is the 
Hubble constant. He speculated that at this point there would be a 'break in the response 
of the vacuum': the waves of Unruh radiation would be unobservable. He further 
speculated that this could have an effect on inertia, if inertia is linked to a form 
of Unruh radiation, as suggested by Haisch~$et~al.$~(1994). He suggested this as a cause
of MOND behaviour. Taking the limiting distance to be twice the Hubble distance (a 
Hubble diameter: $\Theta=2c/H$) we can infer the acceleration at which this break 
would happen for Unruh radiation by rearranging 
(\ref{eq:wiens}) as
    \begin{equation} 
    a = \frac{4 \pi^2 \beta c^2}{\lambda_m}.    \label{eq:critacc}
    \end{equation}
Substituting the following values
$\beta=0.2$,
$c=3 \times 10^8$ m s$^{-1}$ and
$\lambda_m=\Theta=2c/H=2.7 \times 10^{26}$ m
(since $H=2.3 \pm 0.9 \times 10^{-18}$ s$^{-1}$)
the predicted critical acceleration is $a=26 \times 10^{-10}$ m s$^{-2}$. Below this
acceleration inertia could be affected by Milgrom's break. This is larger than 
the acceleration constant of $a=1.2 \times 10^{-10}$ m s$^{-2}$ required for 
MOND (Milgrom~1983) for fitting galaxy velocity curves. It is close to the Pioneer 
anomaly, but Milgrom's (abrupt) break cannot explain the Pioneer anomaly, since the 
Pioneers' acceleration at 50 AU from the Sun was still too large, about 
$10^{-5}$ m s$^{-2}$, and this acceleration implies Unruh wavelengths of 
only 0.03 \% of the Hubble distance.

\subsection{A Casimir-like effect at the Hubble scale}

Here, Milgrom's long-wavelength cutoff idea is modified so that we assume that only 
wavelengths of the Unruh radiation that fit exactly into twice the Hubble distance 
($\Theta=2c/H$) are allowed: those harmonics with nodes at the boundaries. This is 
a similar idea to the Casimir effect in which the energy of the zero-point field 
is reduced between conducting plates because only certain wavelengths can exist 
between them (Casimir,~1948).

Figure~1 shows the energy of Unruh radiation as a function of the wavelength. The 
allowed wavelengths are shown by the dashed vertical lines. As for the Casimir effect, 
these wavelengths are given by
    \begin{equation} 
    \lambda_n = \frac{2 \Theta}{n}.    \label{eq:casimir}
    \end{equation}
where n=1,2,3...etc. For an object with high acceleration the temperature of the 
Unruh radiation is high, the Unruh wavelengths seen are short and the Unruh energy 
spectrum looks like the curve on the left. In the schematic this spectrum is sampled
by five or six of the allowed wavelengths so much of the energy in the Unruh spectrum 
remains. However, if the acceleration is reduced, then the object sees the spectrum 
on the right. In this case, only one of the wavelengths is allowed because the 
others do not fit within $\Theta$ and so the spectrum is more sparsely sampled, 
and the energy of the Unruh radiation is much lower than expected. In this 
new scheme, some spectral energy is lost at wavelengths shorter than $\Theta$, 
and this allows the prediction of the Pioneer anomaly, which cannot be explained 
by the more abrupt $break$ mentioned in Milgrom~(1994,~1999) and discussed in section~2.1. 

If the Unruh energy spectrum is given by a function f($\lambda$), then the unmodified 
inertial mass ($m_i$) is assumed here to be proportional to the integral of this
    \begin{equation} 
    m_i \propto \int_0^\infty f(\lambda) d \lambda. \label{eq:extra1}
    \end{equation}
To model the effect of the increasingly sparse sampling of the spectrum at long wavelengths 
the weight of longer wavelengths in equation (\ref{eq:extra1}) is reduced by using a 
factor $F$ to account for the reduction in sampling density when going from the continous 
sampling of the spectrum to the discrete sampling. By direct calculation it was found
that the number of allowed wavelengths available to sample the Planck spectrum varied 
linearly as $\lambda_m^{-1}$ over the range of wavelengths studied here, where 
$\lambda_m$ is the peak wavelength of the spectrum (this was done by counting the 
number of allowed wavelengths where the spectral energy was more than 1\% of the 
peak energy). Therefore we assume that $F=\frac{A}{\lambda_m} + B$, where A and B are 
constants. When $\lambda_m \rightarrow 0$ the normal continuous sampling 
should be recovered and $F=1$. When $\lambda_m \rightarrow 4 \Theta$ no energy is sampled 
so $F=0$ (this is Milgrom's break, as discussed above). Using these conditions, A and B can
be found and the factor can be shown to be $F=1-{\lambda_m}/4 \Theta$. The model for the 
modified inertial mass ($m_I$) is therefore
    \begin{equation} 
    m_I \propto \int_0^\infty f(\lambda) d \lambda \left( 1-\frac{\lambda_m}{4\Theta} \right). \label{eq:extra2}
    \end{equation}
From equations (\ref{eq:extra1}) and (\ref{eq:extra2})
    \begin{equation} 
    m_I = m_i \left( 1-\frac{\lambda_m}{4\Theta} \right).  \label{eq:modiner2}
    \end{equation}
Using (\ref{eq:wiens}) and assuming the equivalence principle applies to the 
unmodified inertial mass: $m_i=m_g$, the modified inertial mass $m_I$ becomes
    \begin{equation} 
    m_I = m_g \left( 1- \frac{\beta \pi^2 c^2}{a \Theta} \right).  \label{eq:modiner3}
    \end{equation}
Here, $m_I$ behaves in a similar way to what would be expected for MOND (Milgrom,~1983).
For large accelerations the second term in the brackets is negligible and the standard 
inertial mass is recovered. However, as the acceleration decreases, the second term 
becomes larger, and $m_I$ falls further below $m_g$. 
For accelerations much lower than seen here, it is possible for the term in brackets 
to be negative, implying a negative inertial mass. However, in this model, such a 
low acceleration would never be attained, since a body with an inertial mass 
approaching zero would tend to accelerate again: there is a minimum acceleration.
For an acceleration of 9.8~m~s$^{-2}$ 
the inertial mass of a 1~kg object is predicted to be $7 \times 10^{-11}$ kg lower. 
For the small accelerations seen by the Pioneer craft, which are far from a gravitational 
source, the inertial mass is predicted to decrease by 0.01 $\%$. At some point the 
acceleration, acting now on a lower inertial mass, increases again. Eventually a 
balance is achieved, as modelled below, around a particular acceleration.
Assuming modified inertia, the equation of motion for the Pioneer craft is
    \begin{equation}
    F = m_I a = \frac{G M_{\odot} m_{g}}{r^2},   \label{eq:eqofmot1}
    \end{equation}
where $M_{\odot}$ is the solar mass and $r$ is the distance from the Sun. Substituting for $m_I$
from (\ref{eq:modiner3}) we can find the balance point mentioned above
    \begin{equation}
    a = \frac{G M_{\odot}}{r^2} + \frac{\beta \pi^2 c^2}{\Theta}.   \label{eq:eqofmot2}
    \end{equation}
Therefore the acceleration is given by the usual Newtonian inverse square law, but 
with an additional constant term caused by the loss of inertia. This new term has a 
value of $6.9 \pm 3.5 \times 10^{-10}~m~s^{-2}$ which is about six times larger than
the $1.2 \times 10^{-10}~m~s^{-2}$ required for MOND. The 40 \% ($\pm 3.5$) 
uncertainty arises because of uncertainties in the Hubble constant (see section 2.1).

According to (\ref{eq:eqofmot2}) all bodies, even if there is no source of gravity 
($M_{\odot}=0$), would show a minimum acceleration, given by the second term on the 
right hand side, which can be rearranged to give 
$\frac{1}{2} \beta \pi^2 cH \sim 0.99 \times cH$ which is 
close to the observed Hubble expansion rate ($cH$). Therefore
    \begin{equation}
    a = \frac{G M_{\odot}}{r^2} + 0.99 \times cH.   \label{eq:eqofmot3}
    \end{equation}

\section{RESULTS}

The vertical error bars in Figure 2 show the observed Pioneer anomaly as a function of 
distance from the Sun out to 45~AU taken from Anderson~$et~al.$~(2002). Within 
about 10 AU of the Sun the anomaly was indistinguishable from zero. It increased after about 
10 AU to an approximately constant value of 8.74 $\times$ 10$^{-10}$ m~s$^{-2}$.

The solid line shows the acceleration anomaly predicted by the extra term in 
(\ref{eq:eqofmot2}) and the horizontal dashed lines show the error bars for 
the prediction. The predicted anomaly was a constant 
$6.9 \pm 3.5 \times$ 10$^{-10}$ m~s$^{-2}$, 
which is in agreement with the observed anomaly from 10 to 45~AU from the Sun. 

The model predicts that the anomaly should also be found within 10 AU of the Sun and
this does not agree with the first data point at 6~AU from Anderson~$et~al.$~(2002)'s 
data (see the left-most bar on Fig.~2) which shows no anomaly. Also, the planets do 
not show an anomaly. This difference may be due to the Pioneers' unbound trajectory. 
As noted by Milgrom~(2005), for theories of modified inertia the acceleration depends 
on the trajectory as well as the position. A further analysis of the Pioneer data is 
ongoing (Toth and Turyshev,~2006) and should improve the data resolution at the 
crucial point where the Pioneers' trajectories became unbound: between 5 and 10~AU.

The fit of this model to the Pioneer data is less close than that obtained by Moffat 
and Brownstein~(2006). However, they fitted their model to the Pioneer anomaly data 
using two adjustable parameters, whereas there are no adjustable parameters here.

\section{DISCUSSION}

One of the consequences of this idea, not considered by the parameterisation of section
2.2, is that at certain accelerations the Unruh spectral peak is directly sampled by 
the allowed wavelengths, and $m_I/m_g$ is then at a temporary peak. At other accelerations
the nearest sampling wavelength would be slightly off-peak and so $m_I/m_g$ would be lower. 
These ideas therefore predict that the Pioneer data (and also galaxy rotation curves) may 
show a radial variation in the ratio of $m_I/m_g$ as the favoured accelerations are sampled 
one by one moving out from the Newtonian regime near the centre of the Solar System (or 
galaxy) to the lower accelerations further out, with further consequences for dynamics.
At 40 AU from the Sun, the number of allowed wavelengths in the Unruh spectrum seen
by the Pioneers, counted as described in section 2.2, is about 4000. Thus the spectrum is 
still quite well sampled, and these variations may be too small to detect. However, near the 
edges of galaxies, accelerations are much lower, and the Unruh spectrum would be 
sampled by only a few wavelengths. Therefore, the differences in the ratio $m_I/m_g$ 
between a case in which the discrete sampling hits the spectral peak, and a case in which 
it misses it, would be more obvious, and the impact on stellar dynamics of the variations 
should be more easily detected.

As mentioned above, an analysis of newly recovered Pioneer data from the inner solar 
system is currently in progress (Toth and Turyshev,~2006) and would support a theory 
of modified inertia, though not necessarily this one, if it is confirmed that the 
anomaly began at the same time that the Pioneer probes moved from bound orbits to 
hyperbolic ones (Milgrom~1999). These new data may also resolve the direction of the 
anomalous force. An acceleration towards the Sun would imply modified $G$, one towards 
the Earth would imply a problem with time, and an acceleration along the Pioneer 
trajectory would imply some kind of modified inertia.

Zhao~(2005) and Zhao and Tian~(2006) have shown that if MOND is true instead of Newtonian 
theory, then Roche lobes should be more squashed and therefore it should be possible to
test for MOND by investigating a local Roche lobe. This test could also differentiate 
between modified gravity and modified inertia versions of MOND, since for modified 
inertia the shape of the Roche lobe would depend on the approach trajectory of 
the probe, and for modified gravity it would not.

In this scheme there is a minimum allowed acceleration which depends on a Hubble
scale $\Theta$, so, if $\Theta$ has increased in cosmic time, there should be a 
positive correlation between the anomalous centripetal acceleration seen in equivalent 
galaxies, and their distance from us, since the more distant ones are seen further 
back in time when, if the universe has indeed been expanding, $\Theta$ was smaller. 
The mass to light ratio (M/L) does seem to increase as we look further away. The 
M/L ratio of the Sun is 1 by definition, for nearby stars it is 2, for galaxies' it 
is 50, for galaxy pairs it is 100 and for clusters it is 300.
As an aside: equation (\ref{eq:eqofmot2}) could be used to model inflation, since 
when $\Theta$ was small in the early universe the minimum acceleration is predicted 
to be larger.

Part of this scheme is the hypothesis that Unruh radiation of very low temperature 
is weaker than expected, because of a wavelength limit, so it is logical to extend 
this to the temperature of any object. If the limiting wavelength idea is correct, 
then the energy radiated by a very cold object should be less than that expected 
from the Stefan-Boltzmann law. The coldest temperature achieved so far is 100~pK 
at the Helsinki University of Technology (Knuuttila,~2000). Using Wien's law, an 
object this cold would have a peak radiating wavelength of $3 \times 10^7~m$. 
By analogy to (\ref{eq:modiner2}) the energy of the black body radiation spectrum 
($E$) would be modified to $E'$ as
    \begin{equation} 
    E' = E \left( 1-\frac{\lambda_m}{4\Theta} \right) = E (1 - 2.7 \times 10^{-20}) J. \label{eq:bh1}
    \end{equation}
It is unknown to the author whether differences in radiating energy as small as this can be 
detected.

The Hawking (1974) temperature of a black hole is given by a very similar expression to
that of Unruh but involving the mass of a black hole $M$:
    \begin{equation} 
    T=\frac{\hbar c^3}{8 \pi G M k}.   \label{eq:bh2}
    \end{equation}
As in section 2.1, we can use Wien's law ($T=W/{\lambda_m}=\beta h c / k \lambda_m$) 
again to substitute for $T$ and impose a limit on the allowed wavelength
    \begin{equation} 
    \frac{16 \pi^2 G M \beta}{c^2} \leq \frac{2c}{H}.    \label{eq:bh3}
    \end{equation}
Therefore
    \begin{equation} 
    M \leq \frac{c^3}{8 \pi^2 G \beta H}.    \label{eq:bh4}
    \end{equation}
Substituting values as follows:
$c=3 \times 10^8$ m s$^{-1}$,
$G=6.67 \times 10^{-11}$ Nm$^2$ kg$^{-2}$,
$\beta=0.2$,
$H=2.3 \times 10^{-18}$ s$^{-1}$ we get $M \leq 1 \times 10^{52}$ kg. This is a predicted 
maximum mass of a black hole: about $10^{22}$ solar masses.

The assumptions made in equation (\ref{eq:extra1}) and (\ref{eq:extra2}) have not 
individually been verified, but they do produce results similar to the Pioneer anomaly. 
A criticism of this scheme could be that the parameterisation of the decrease in sampling 
density neglects subtle variations as the Unruh spectrum falls between allowed 
wavelengths, and these variations could be useful for testing the idea. The 
simple model developed here should ideally be replaced by a model that calculates $m_I$ 
more directly, by sampling the Unruh spectrum discretely.

\section*{CONCLUSIONS}

Two hypotheses were made: (1) Inertia is a reaction to Unruh radiation and 
(2) this reaction is weaker for low accelerations because some wavelengths 
in the Unruh spectrum do not fit within a limiting scale (twice the Hubble 
distance) and are disallowed: a process similar to the Casimir effect.

Using these ideas, the Pioneer acceleration anomaly was predicted to be 
$6.9 \pm 3.5 \times 10^{-10}$ m s$^{-2}$, which agrees within 
error bars, beyond 10~AU from the Sun, with the observed value of 
$8.74 \pm 1.33 \times 10^{-10}$ m s$^{-2}$.

This scheme is appealingly simple, and does not require adjustable parameters. 
However, the model predicts an anomaly within 10 AU of the Sun which is not 
observed. Various tests of this idea are also discussed, including the 
possibility that subtle variations in galaxy acceleration curves 
(if not the Pioneer data) might be detectable.

\section*{Acknowledgements}

Many thanks to M.Milgrom and C.Smith for commenting on earlier drafts, 
to an anonymous reviewer who made many useful suggestions, 
and to B.~Kim for support and encouragement.

\section*{References}

Anderson,~J.D., P.A.~Laing, E.L.~Lau, A.S.~Liu, M.M.~Nieto and S.G.~Turyshev, 1998.
Indication, from Pioneer 10/11, Galileo and Ulysses Data, of an apparent weak
anomalous, long-range acceleration. $Phys.~Rev.~Lett.$, 81, 2858-2861.

Anderson,~J.D., P.A.~Laing, E.L.~Lau, A.S.~Liu, M.M.~Nieto and S.G.~Turyshev, 2002.
Study of the anomalous acceleration of Pioneer 10 and 11. preprint, arXiv:gr-qc/0104065v5

Baessler,~S, B.R.~Heckel, E.G.~Adelberger, J.H.~Gundlach, U.~Schmidt and H.E.~Swanson, 
1999. Improved Test of the Equivalence Principle for Gravitational Self-Energy.
$Phys.~Rev.~Lett.$, 83 (18), 3585.

Bekenstein,~J.D.,~2004. Relativistic gravitation theory for the MOND paradigm.
$Phys.~Rev.~D$, 70, 3509.

Casimir,~H.B.G.,~1948. $Proc.~Kon.~Nederland.~Akad.~Wtensch.$, B51, 793.

Eisberg,~R, and R.~Resnick,~1985. Quantum Physics (2nd Ed.). John Wiley and Sons Ltd.

Haisch,~B., Rueda,~A. and Puthoff,~H.E., 1994. Inertia as a zero-point field Lorentz
force. $Phys.~Rev.~A$,~49,~678.

Hawking,~S.,~1974. Black hole explosions. $Nature$, 248, 30.

Knuuttila,~T.A.~2000, Nuclear magnetism and superconductivity in rhodium. DSc thesis.
Helsinki Univeristy of Technology. \\ http://lib.tkk.fi/Diss/2000/isbn9512252147/

Mannheim, P.D.,~1990. Conformal cosmology with no cosmological constant,
$General~Relativity~and~Gravitation$, 22, 289.

Milgrom~M.,~1983. A modification of the Newtonian dynamics as a possible alternative 
to the hidden mass hypothesis. $Astrophysical~Journal$, 270, 365.

Milgrom~M.,~1994. $Ann.~Phys.$, 229, 384.

Milgrom~M.,~1999. The Modified Dynamics as a vacuum effect. $Phys~Lett~A$, 253, 273.

Milgrom~M.,~2005. MOND as modified inertia. in Mass profiles and shapes of cosmological 
structures. G.Mamon, F.Combes, C.Deffayet and B.Fort (eds). EAS Publications Series, Vol. 9.

Moffat,~J.W. and J.R.~Brownstein, 2006. Gravitational solution to the Pioneer 10/11 anomaly.
$Classical~and~Quantum~Gravity$, 23, 3427-3436. preprint: arXiv:gr-qc/0511026, 17th May 2006.

Sanders, R.H. and S.S.~McGaugh, 2002. Modified Newtonian Dynamics as an alternative to
dark matter. $Annu.~Rev.~Astron.~Astrophys.$, 40, 263-317, 2002.

Sellwood,~J.A.,~2004. What is the evidence for dark matter? Dark Matter in Galaxies, 
IAU Symposium., 220. preprint: arXiv:astro-ph/0401398.

Toth,~V.T. and S.G.~Turyshev,~2006. The Pioneer anomaly: seeking an explanation in newly 
recovered data. preprint: arXiv:gr-qc/0603016v1., 7th March 2006.

Unruh,~W.G.,~1976. Notes on black hole evaporation, $Phys.~Rev.~D$, 14, 870.

Van Allen,~J.A.,~2003. Gravitational assist in celestial mechanics - a tutorial.
$Am.~J.~Phys.$, 71 (5), 448-451.

Zhao,~H.,~2005. Roche lobe sizes in deep-MOND gravity. $Astronomy$ $and$ $Astrophysics$, submitted.
preprint: arXiv/astro-ph/0511713

Zhao,~H.,~B.~Famaey,~2006. $ApJ~Letters$. In press, preprint: arXiv/astro-ph/0512425

Zhao,~H, and L.~Tian,~2006. Roche lobe shapes for testing MOND-like modified gravities.
$Astronomy~and~Astrophysics$, 450 (3), 1005-1012.

Zwicky,~F.,~1933. $Helv.~Phys.Acta$,~6,~110.

\newpage

\begin{figure}
\begin{center}
\resizebox{300pt}{300pt}{ \includegraphics{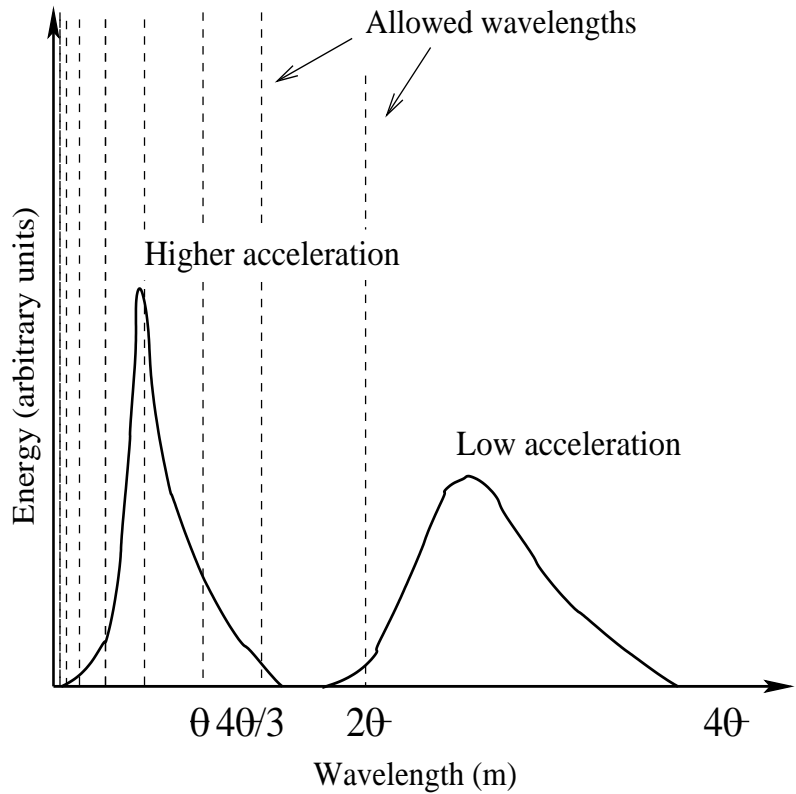}}

\vspace{1cm}

\caption{A schematic in which the vertical dashed lines show the wavelengths that 
fit within twice the Hubble distance and are allowed in this model. Unruh spectra 
for different accelerations are also shown. The one on the right represents a 
lower acceleration, and is more sparsely sampled by the allowed wavelengths.}
\end{center}
\end{figure}

\newpage

\begin{figure}
\resizebox{300pt}{300pt}{ \includegraphics{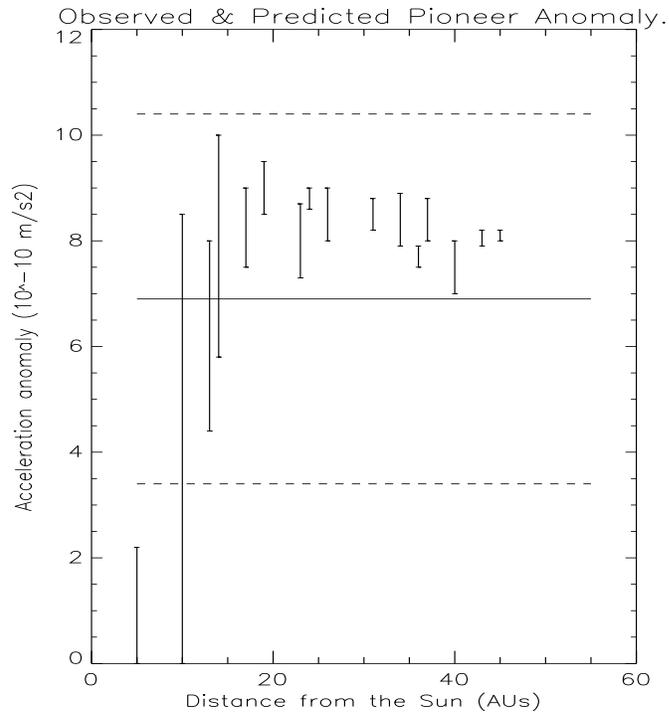}}

\vspace{1cm}

\caption{The bars show the observed Pioneer 10 and 11 anomalies as a function 
of distance from the Sun (AU) (taken from Anderson~$et~al.$,~2002). The solid 
line shows the Pioneer anomaly predicted by equation \ref{eq:eqofmot2} and the
dashed lines represent the errors bars for the model.}
\end{figure}

\end{document}